\documentstyle[twocolumn,aps,psfig]{revtex}
\begin{document}
\title{Periodic orbit effects on conductance peak heights in a chaotic
quantum dot}
\author{L. Kaplan\thanks{lkaplan@phys.washington.edu}
 \\Department of Physics and Institute for Nuclear Theory,\\
 University of Washington, Seattle, WA 98195}
\maketitle

\begin{abstract}
We study the effects of short-time classical dynamics on the distribution
of Coulomb blockade peak heights in a chaotic quantum dot. The
location of one or both leads relative to the short unstable orbits, as
well as relative to the symmetry lines, can have large effects on the moments
and on the head and tail of the conductance distribution.
We study these effects
analytically as a function of the stability exponent of the orbits
involved, and also numerically using the stadium billiard as a model.
The predicted behavior is robust, depending only on the short-time behavior
of the many-body quantum system, and consequently insensitive
to moderate-sized perturbations.
\end{abstract}

\vskip 0.1in

\section{Introduction}

Quantum dots, semiconductor devices in which electrons
are confined to live inside a two-dimensional mesoscopic-sized
region, have generated much experimental and theoretical interest
in the past decade~\cite{quandot}.
In the Coulomb blockade regime~\cite{coulblock}, the dot is weakly coupled
to the outside through two narrow or tunneling leads, and individual
resonances can be observed when the Fermi energy in the leads
matches the energy
of a state of $N$ electrons in the dot. As a function of the Fermi energy
or gate voltage one then observes a series of peaks in the conductance,
the peak width being controlled by the temperature and the spacing 
between them by the classical
charging energy required to add one more electron to the dot (in the
experimentally typical regime where the level spacing is large
and the intrinsic resonance width small compared with the temperature).
The conductance peak height of the $n$-th
resonance is then given by
\begin{equation}
G_n = {e^2 \over h} {\pi \over 2 kT} \, g_n\,,
\end{equation}
where
\begin{equation}
\label{defgn}
g_n = {\Gamma_{an} \Gamma_{bn} \over \Gamma_{an} + \Gamma_{bn}}\,,
\end{equation}
and $\Gamma_{an}$, $\Gamma_{bn}$ are the partial decay widths
of the $n$-th resonance through each of the two leads labeled
$a$ and $b$.

Each of the two partial widths is given by Fermi's Golden Rule
as the square of a tunneling matrix element. This matrix element
in turn is obtained (in the single-particle picture)
by taking the overlap of the normal derivative
of the $n$-th dot wavefunction along the boundary with the electron
wavefunction
in the lead~\cite{narim,matrelem}:
\begin{equation}
\Gamma_{an} = C_a \left| \langle \partial_\perp\Psi_n | \phi_a\rangle
\right|^2 
\label{gammapsi}
\end{equation}
(and similarly for $\Gamma_{bn}$),
where $\phi_a$ is the relevant transverse wavefunction
in lead $a$, $\partial_\perp\Psi_n$ is the normal derivative of the $n$-th
dot eigenstate at the boundary, and $C_a$ is a constant associated
with the height of the barrier (possibly $C_a \ne C_b$ if the two leads
are unequally coupled).
Because tunneling will always be
dominated by the lead mode which has the largest longitudinal energy,
we may without loss of generality
restrict ourselves to $\phi_a$ which is the lowest transverse
energy mode of the lead~\cite{narim}. For a smooth lead potential,
this will be given by a Gaussian,
\begin{equation}
\phi_a \sim e^{-(q-q_0)^2/2\sigma^2}\,,
\label{leadmode}
\end{equation}
where the width $\sigma \sim
\sqrt \hbar$ depends on the detailed properties of the lead.

Of course in reality the multi-electron state inside the dot is not
given by a product of single-particle states, nor do we know the electronic
Hamiltonian inside the dot well enough to have any realistic hope of being
able to compute the wavefunctions $\Psi_n$. We will come back to these
important issues in the next section.

Many authors~\cite{rmtcond}
have studied the behavior of the conductance peaks $G_n$
in the context of random matrix theory. There the overlaps
$\langle \partial_\perp \Psi_n | \phi_a\rangle$ are considered
as random Gaussian variables (real or complex), and thus the widths
$\Gamma_{an}$, $\Gamma_{bn}$ become $\chi^2$ random variables of
one or two degrees of freedom in the absence or presence of
a magnetic field, respectively. These predictions have compared
favorably with the experimental data~\cite{exper}.
In the present work, we extend these dynamics-free
results to include the effects of short-time dynamics on the distribution
of conductance peak heights through a ballistic dot. In doing so,
we are following the work of Narimanov, Cerruti, Baranger, and
Tomsovic~\cite{narim} who have already treated the special case of two
leads placed symmetrically on the horizontal-bounce orbit
of a stadium billiard (though
focusing on peak-to-peak correlations, rather than
on the peak heights themselves, in contrast with the present work).
Here we consider in full generality
the short-time classical
effects on conductance peak heights, including the dependence
of the peak distribution on the stability exponents of the
short orbit or orbits near which one or both leads may be
located. We also disentangle the effects of symmetry lines
and symmetric lead placement from the effects of short-time
classical dynamics.

We note that although quantum dot experiments provide one of 
the motivations for the present work, the effects we are considering
are relevant to a wide variety of physical situations where observable
properties are affected by the statistical properties of wavefunctions.
In particular, the conductance problem discussed here is formally
analogous to a situation often encountered in molecular or nuclear
physics, where a reaction $A \to B \to C$ takes place through
an intermediate or transition state $B$ living inside
a metastable ``well," and the reaction rate is then determined 
by the structure of the wavefunction $B$ inside the well.

The paper is organized as follows: in Section~\ref{secscars} we briefly
review some recent results concerning short-time dynamical effects
on wavefunction intensities. We focus in particular on the separation
of scales between the bounce time in the dot and the time
at which eigenstates are resolved, and on the consequent robustness
of short-time effects on {\it statistical} wavefunction properties.
In Section~\ref{seccond} we analyze the effects of short periodic
orbits (s.p.o.'s) on conductance peak statistics for several qualitatively
different situations: one lead on a s.p.o., two leads on the
same s.p.o., and two leads on different s.p.o.'s for which the
spectral envelopes may be in or out of phase with each other
in the energy range
of interest. Numerical tests of these predictions appear
in Section~\ref{secnumtests}, followed by concluding remarks
in Section~\ref{secconc}.

\section{Scars and wavefunction intensities: basic results}
\label{secscars}

The scar effect is one of the most visually striking aspects
of quantum chaotic behavior. It was noticed already in the
1980's that the quantum wavefunctions of classically
chaotic systems display anomalous enhancement and suppression
of intensity in the vicinity of the unstable periodic orbits,
contrary to the naive expectation of wavefunction randomness
and uniformity~\cite{mcdonald}. The early theories of this
phenomenon~\cite{earlyscar} treated the short-time
linearized dynamics around an unstable orbit, and thus made
predictions about energy-smoothed spectral properties. More
recently, theories of scars were extended to include long-time
nonlinear recurrences, making possible predictions about the
distribution of individual eigenstate intensities on a given
periodic orbit~\cite{recentscar}. The scar formalism has also
been adapted to study quantitatively wavefunction
structure in systems as varied as Sinai-type billiards
and in two-body random interaction ensembles in nuclear
physics.

The key result of this work that
we need for the present analysis is that wavefunction intensities
in a closed system are given by
\begin{equation}
|\langle \Psi_n | \phi \rangle|^2 =
 r_n S^{\rm smooth}_{\phi}(E_n) \,,
\label{intfluct}
\end{equation}
where $S^{\rm smooth}_{\phi}(E)$ is a smooth
local density of states appropriate to the test state
$\phi$, and the $r_n$ are random $\chi^2$ variables.
The smooth envelope can be determined by Fourier transforming
the short time autocorrelation function of
$\phi$:
\begin{equation}
S^{\rm smooth}_{\phi}(E)={\rm FT}
[A^{\rm short}_{\phi}(t)] \,,
\label{ft}
\end{equation}
where 
\begin{equation}
A^{\rm short}_{\phi}(t)= 
\langle \phi|\phi(t) \rangle e^{-t^2/T_0^2}
\end{equation}
with some cutoff time $T_0$. Now as long as the cutoff $T_0$
is chosen to be small compared with the Ehrenfest time, which scales
as the inverse Lyapunov exponent of the system times
$\log{(k_FL)}$ ($L$ being the size of the dot), the short-time
dynamics of the wavepacket $\phi$ may be determined
semiclassically, using classical motion linearized around the
center of the wavepacket. 

In particular, let $\phi$ (which we will later identify with
$\phi_a$) be a Gaussian wavepacket centered on an unstable periodic
orbit of period $P$. Then we have for $t=mP$~\cite{antiscar}
\begin{equation}
\label{aphi}
A_{\phi}^{\rm short}(t)={e^{-im\theta} \over
\sqrt{\cosh \beta m + i Q \sinh \beta m}}\,,
\end{equation}
where $\beta > 0$ is the instability exponent for one iteration
of the periodic orbit, $-\theta$ is the classical action in units
of $\hbar$ (plus Maslov indices, if any), and $Q$
is a non-optimality parameter. The largest recurrences ($Q=0$)
are obtained when the initial wavepacket $\phi$ is optimally oriented with
respect to the stable and unstable manifolds of the orbit. For a lead
located on a fixed periodic orbit, $Q$
will in general be a function of the width $\sigma$ of the Gaussian
mode in the lead (see Eq.~\ref{leadmode} and
Ref.~\cite{antiscar}). More
important than the analytic form of Eq.~\ref{aphi},
however, is the fact that for weakly
unstable orbits (strictly $\beta \ll 1$, though due to numerical
factors $\beta \approx 2$ is already in a sense ``weak"),
strong recurrences in $A(t)$ persist for $O(\beta^{-1})$ periods. In the
energy domain, we get bumps in the smoothed local density of states
$S^{\rm smooth}_\phi(E)$, of width $O(\beta) \ll 1$ compared to the spacing
between the bumps, and of height $O(\beta^{-1}) \gg 1$ compared to the
mean. Very roughly speaking then, $O(\beta)$ of all wavefunctions are
``scarred" on the periodic orbit, having intensity there
$O(\beta^{-1})$ greater than the mean, while most of the remaining
wavefunctions are ``antiscarred," their intensity on the periodic orbit
being much smaller than the mean. This separation of wavefunctions
into scarred and antiscarred ones is of course only a oversimplified
picture and one can obtain quantitatively
the full distribution of wavefunction
intensities on the orbit as a function of the exponent $\beta$;
such a comparison between an analytic result for the tail of the intensity
distribution and numerical data was in fact performed in~\cite{wis}.
One can also study the moments of the intensity distribution, and finds,
for example, that the mean squared intensity approaches $\pi /\beta$
times the RMT expectation for small $\beta$ (in units where the mean
intensity in normalized to one).

We emphasize the distinction between scar predictions and the brute-force
semiclassical computation of chaotic wavefunctions~\cite{ltsc}. In many
physically interesting situations, the Hamiltonian of the system is not known
nearly well enough to compute individual eigenlevels and eigenstates
either semiclassically or indeed using the full quantum machinery. What is
of interest in such situations is not so much
the detailed structure of the
$n$-th eigenstate in a given sample, but rather attaining
a theoretical understanding of the statistical properties of the system.
In a billiard (hard wall) system, changing the boundary by even one square
wavelength far away from the periodic orbit of interest will already
destroy the detailed structure of individual wavefunctions on the orbit,
but will not affect statistical properties such as the distribution of
intensities on the orbit. In the time domain we see this easily by
recognizing that to resolve individual eigenstates requires following
the dynamics to times of order of the Heisenberg time, which scales as
$\hbar$ over the mean level spacing, or $O(k_FL)$ bounce times
in $d=2$. Scar predictions, on the other hand,
only require short-time dynamical information, on the scale of the one-bounce
time $T_B$ (or, to get the full effect, the classical Lyapunov decay
time $T_B/\beta$ if the instability exponent
$\beta$ is small). Thus, deformations of the Hamiltonian will not
affect these statistical predictions as long as the mean free path
associated with such deformations is larger than the sample size
$L$ (i.e. as long as we truly remain in the ballistic regime).

In the energy domain, the mean level spacing in $d=2$ goes as $1/L^2$
(in units where $m,\hbar \sim 1$), while the scale associated with scar effects
(i.e. the separation between the bumps in the smooth
local density of states envelope)
is $1/T_B \sim k_F/L$. Thus a single bump in the LDOS corresponds to
$O(k_FL) \gg 1$ level spacings, and in the high-energy regime (i.e. many
electrons in the dot) the
resulting wavefunction intensity statistics will be quite insensitive
even to perturbations that are large compared with the mean level
spacing.

The preceding argument applies also to electron-electron
scattering effects. It is known that a rather weak interaction
beyond mean field
will completely destroy level repulsion and
cause the distribution of level spacings to approach a
Gaussian form, in strong contrast with the Wigner-Dyson prediction
of the single-particle theory~\cite{levelspac}. The Gaussian
behavior is indeed what is observed experimentally in such systems.
This is not surprising because strong multi-particle effects on the
level spacing scale require only that the interaction mean free
path be smaller than the very large
Heisenberg scale $\sim (k_FL) T_B$. On the other hand, as we have seen,
the effects on wavefunction statistics will be weak as long as
an individual electron can freely travel across the device before
interacting.

It is also known (by comparing the ground state of a dot with $N$
electrons with the excited states of the same dot containing only
$N'<N$ electrons)
that adding electrons to the dot changes the shape of the mean field potential
and thus has a significant effect on the character of the single-particle
states. Arguments very similar to those in the preceding paragraphs tell
us that a very small change
in the effective potential (resulting in matrix elements of the perturbation
which are of order of the mean level spacing)
is sufficient to destroy our predictive power
for individual wavefunctions, but does not affect the {\it statistical}
properties
of these wavefunctions, which are associated with a much shorter time scale
(the bounce time)
and are therefore robust to any such perturbation. The statistical predictions
would become irrelevant only if adding one or a few electrons to the dot
changed the resulting potential in such a way as to completely change
the character of the short classical trajectories.

\section{Conductance peak heights in chaotic systems}
\label{seccond}

\subsection{Generically placed leads}

Within the context of RMT,
Alhassid and Lewenkopf~\cite{alh} have derived
an explicit form for the distribution of conductance peak heights
$g_n$, allowing for the possibility of unequal leads ($C_a \ne C_b$
above) and also allowing for correlated channels. In the presence
of time reversal symmetry,\footnote{For definiteness, we will consider
throughout the time-reversal invariant situation. Of course, the calculations
can be carried through also in the presence of a magnetic field
(the wavefunction intensity distribution for that case has been studied
extensively in\cite{wis,antiscar}),
and all predicted effects are qualitatively similar there.}
for two leads consisting of one channel each,
this distribution reduces to
\begin{eqnarray}
P(g) & = &
\int_{1 / 2s_a}^\infty d\tau \int_{1 / 2s_b}^\infty d\tau'
{\left(4g/2\pi^2\right)
e^{-(\tau+\tau')g} \over \sqrt{(s_a\tau-1/2)(s_b\tau'-1/2)}} \nonumber
\\ & \times &\left[ K_0 \left(2g \sqrt{\tau \tau'}\right) + {1 \over 2}\left(
\sqrt{\tau \over \tau'}+\sqrt{\tau' \over \tau} \right)
K_1 \left(2g \sqrt{\tau \tau'}\right)\right] \nonumber \\ & = &
{1 \over \sqrt {2 \pi g}} \left ( {1 \over \sqrt{s_a}}+{1 \over \sqrt{s_b}}
\right) \exp\left[-{g \over 2} \left ( 
{1 \over \sqrt{s_a}}+{1 \over \sqrt{s_b}}
\right)^2 \right ] \nonumber \\ & = &
\sqrt {2  \over \pi s_\ast g} e^{-2g/s_\ast} \,,
\label{general}
\end{eqnarray}
where $s_a$ and $s_b$ are the mean partial widths through the two
leads $a$ and $b$, and in the last line we use the fact that the distribution
depends only on the quantity $s_\ast$ defined by
\begin{equation}
{1 \over \sqrt{s_\ast}} = {1 \over 2} \left( {1 \over \sqrt{s_a}} +
{1 \over \sqrt{s_b}} \right) \,.
\label{sast}
\end{equation}
For equal leads, $s_\ast=s_a=s_b$, and we also notice that in general
the distribution of $g$ is just the Porter--Thomas $\chi^2$ distribution
of one degree of freedom, with mean height $s_\ast/4$.

In this paper we will primarily be interested in the physical
case of equal-sized
leads (but see also Section~\ref{secunequal}),
however as we will see below the more general expression is a very
useful starting point for studying periodic orbit and symmetry effects.

For two generically placed equal-sized leads, we have $s_a=s_b=s_0$,
and 
\begin{equation}
P_{\rm generic}(g) =\sqrt {2  \over \pi s_0 g} e^{-2g/s_0}\,.
\label{generic}
\end{equation}
The properties of this Porter--Thomas distribution are well-known;
in particular 
\begin{equation}
\langle g\rangle_{\rm generic} = {s_0 \over 4} \;\;\; {\rm and}
\;\;\; \langle g^2 \rangle_{\rm generic} = 3
\langle g\rangle_{\rm generic}^2 \,.
\label{genmoms}
\end{equation}
The ratio of the mean squared height to the square of the mean,
also known as the inverse participation ratio (IPR), is the
simplest measure of the degree of fluctuation in peak
intensities.

We note also that for two leads that are symmetrically placed in 
a dot with reflection symmetry, the two partial widths
$\Gamma_{an}$ and $\Gamma_{bn}$ are equal for each resonance $n$,
and we have $g_n={\Gamma_{an}}/2$. The peak heights are then again
distributed according to a Porter--Thomas law, but with a larger
mean height:
\begin{eqnarray}
P_{\rm symmetric}(g) &=&\sqrt {1  \over \pi s_0 g} e^{-g/s_0} \nonumber \\
\langle g\rangle_{\rm symmetric} &=& {s_0 \over 2}  \,.
\label{symmetric}
\end{eqnarray}
This factor of $2$ enhancement associated with perfect correlation
between the two leads, and independent of any short-time dynamical
effects, is present already in the case
of the horizontal bounce orbit considered in Ref.~\cite{narim}. We note,
however, that the symmetry effect is much less robust than the
dynamical effect, requiring the dot potential to be perfectly symmetric
to better than a single level spacing (otherwise the even and odd states mix
and we recover the generic result of Eq.~\ref{generic}).

Another independent effect, also present in the special example treated
in~\cite{narim}, is associated with the placement of one or both
leads on the symmetry lines of the system. If both leads are placed
on a given symmetry line, as in the case of the horizontal bounce
orbit of the stadium, then only the even states produce resonances, leading
to half as many peaks as naively expected, but with double the
mean height of Eq.~\ref{symmetric}:
\begin{equation}
\langle g \rangle_{\rm same \; sym \; line}
 =s_0 \;\;\; [{\rm half \; expected \; density}]\,.
\end{equation}
In a system with two symmetry lines
such as the Bunimovich stadium or the Sinai billiard, 
we may also consider the case of two leads on different symmetry lines.
Then only a quarter of the eigenstates produce conductance peaks, the
two partial widths are uncorrelated but each is doubled with respect
to the naive expectation $s_0$, and we obtain (cf. Eq.~\ref{genmoms})
\begin{equation}
\langle g \rangle_{\rm diff \; sym \; lines} ={s_0 \over 2} \;\;\;
 [{\rm quarter \; expected \; density}]\,.
\end{equation}
Finally, for just one lead on a symmetry line, only the even wavefunctions
contribute with mean partial width $2s_0$ through the symmetry line lead
and $s_0$ through the other lead; the general expression of Eq.~\ref{general}
then leads to
\begin{eqnarray}
\langle g \rangle_{\rm one \; on \; sym \; line}& =&{2 \over (\sqrt 2+1)^2} s_0
\approx 0.343 s_0 \nonumber \\
& &{\rm [half \;  expected \; density]}\,.
\end{eqnarray}
Thus we see that a rich diversity of conductance
behavior may be observed simply
by considering the placement of one or both leads with respect to the 
symmetry lines of the system.

\subsection{One lead on short periodic orbit}
\label{seconelead}

We now put aside symmetry considerations and consider a scenario where
one of the two conducting leads, say $a$, is located on a short (unstable)
periodic orbit of the chaotic dot. Then the mean partial
width through lead $a$ is given by $s_a=s_{a0}S^{\rm smooth}(E)$ (see
Eqs.~\ref{gammapsi},~\ref{intfluct}) in an energy range near $E$.
Taking the two leads to be of equal width, $s_{a0}=s_b=s_0$,
the effective coupling $s_\ast$ becomes energy-dependent:
\begin{equation}
s_\ast(E)=4s_0\left( {1 + 1/\sqrt{S^{\rm smooth}(E)}} \right)^{-2} \,.
\label{sast1}
\end{equation}
Notice that the conductance is always dominated by the more weakly coupled
lead: thus near the scarring energies $S^{\rm smooth}$ is strongly enhanced
and the mean conductance is moderately increased (at most by a factor of $4$),
while at the antiscarring
energies $S^{\rm smooth}$ is small, and the mean conductance may be greatly
suppressed. Sweeping through energy, one obtains a general
expression for the distribution of conductance peak heights:
\begin{equation}
\label{ponelead}
P(g)={1 \over 2\pi} \int_0^{2 \pi} dE \sqrt{2 \over \pi s_\ast(E) g}
e^{-2g/s_\ast(E)}\,,
\end{equation}
where $s_\ast(E)$ is given by Eq.~\ref{sast1} and $S^{\rm smooth}(E)$
by the Fourier transform of Eq.~\ref{aphi}.
Note that the energy interval corresponding
to one oscillation of the scarring envelope is taken as $0$ to
$2 \pi$, because we are working in units where $\hbar=1$ and
the orbit period $P$ is also normalized to $1$. The distribution 
of Eq.~\ref{ponelead} may be computed numerically
(see, for example, Fig.~\ref{figstad} below)
for various
values of the stability exponent $\beta$ (and lead non-optimality
parameter $Q$, see Eq.~\ref{aphi}). First, however, we obtain some
analytic asymptotic results for the strongly scarred case ($\beta \ll 1$),
where
the deviations from Porter--Thomas behavior are expected to be strongest.

The tail of the distribution $P(g)$ will be dominated (for any $\beta$)
by the peak of the envelope $s_\ast(E)$, which coincides of course
with the peak in $S^{\rm smooth}(E)$ at
$E=\theta \; {\rm mod} \; 2\pi$ 
(see Eq.~\ref{aphi}). The integral may be performed
by stationary phase (as in Ref.~\cite{wis}), to obtain
\begin{equation}
\label{asympt}
P(g)={1 \over 2 \pi s_0 g}\sqrt{A \over B} e^{-g A/2s_0 }\,,
\end{equation}
where $A=4/s_\ast^{\rm max}$ and $B$ measures the curvature of the envelope
at the maximum: $B=2 \partial^2(s_\ast^{-1})/\partial E^2$.
Eq.~\ref{asympt} describes the large $g$ behavior of the
conductance peak distribution, where $A$ and $B$ are appropriate functions
of the instability exponent $\beta$. For small $\beta$ one may
simplify further and obtain $A=1 +2 \sqrt{\beta /C}+O(\beta)$ and 
$B={1 \over 2} DC^{-3/2} \beta ^{-3/2}+O(\beta^{-2})$,
where $C \approx 5.24$
and $D \approx 45.1$ are numerical constants. So finally we obtain
the large-$g$ behavior for small $\beta$:
\begin{equation}
\label{oneleadasympt}
P(g)={1 \over \sqrt{2}\pi} {C^{3/4} \over D^{1/2}} {\beta^{3/4} \over g}
e^{-g(1+2\sqrt{\beta/C})/2s_0}\,.
\end{equation}
Notice the long tail dominated by the
exponential behavior $\exp(-g/2s_0)$ in the strongly-scarred
regime, to be contrasted with the much shorter tail $\exp(-2g/s_0)$
in the RMT ($\beta \to \infty$) case.

\begin{figure}
\psfig{file=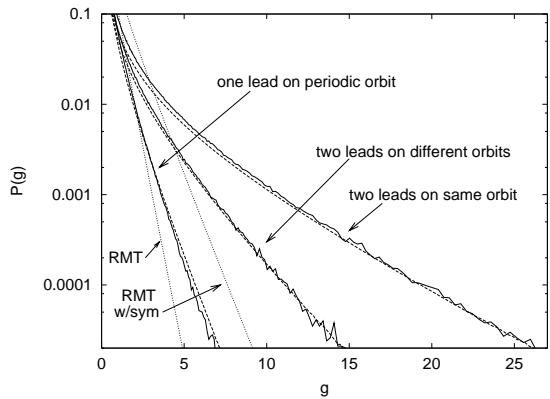,angle=270,width=3in}
\vskip 0.1in
\caption{The tail of the conductance peak height distribution $P(g)$ is plotted
with solid curves for (i) one lead on a periodic orbit with instability
exponent $\beta=1$, with the other lead generically placed; (ii) two leads
on different orbits, each with instability exponent $\beta=1$, and with
in-phase spectral envelopes; and (iii)
two leads placed
on the same orbit with $\beta=1$. Theoretical predictions, given by
Eqs.~\ref{oneleadasympt},~\ref{largeginphaseorbit}, and
~\ref{largegsameorbit}, respectively, and strictly valid for 
$g \gg \beta^{-1} \gg 1$,
are plotted as dashed curves. The mean partial width $s_0$ through each of the 
two leads has been set to unity. For reference, the Porter-Thomas prediction
of RMT (Eq.~\ref{generic},
to be compared with data sets (i) and (ii)) appears as the lower dotted curve,
while RMT modified to include symmetry effects (Eq.~\ref{symmetric}, to
be compared with data set (iii)), is shown as the upper dotted curve. In the 
dynamics-free limit where $\beta \to \infty$, the data would approach these
RMT results.
}
\label{figtail}
\end{figure}

The asymptotic behavior
of Eq.~\ref{oneleadasympt} is plotted for instability 
exponent $\beta=1$ as the leftmost
dashed curve in Fig.~\ref{figtail}, and agrees well with
numerical data, represented by a solid curve. Thus, we see that the 
approximations used in obtaining Eq.~\ref{oneleadasympt} are good already
for $\beta=1$, even though formally we have used a small-$\beta$
approximation. The Porter-Thomas prediction of RMT (Eq.~\ref{generic})
appears in Fig.~\ref{figtail} as a dotted line for comparison.

The predicted increase in the frequency of very small conductance
peaks is even more striking. In Ref.~\cite{antiscar} it was found that
in between the scarring energies
$E=\theta \; {\rm mod} \; 2\pi$ where the LDOS 
$S^{\rm smooth}(E)$ is maximized
and the  antiscarring energies $E=\theta+\pi \; {\rm mod} \; 2\pi$
where it is minimized, the LDOS at the lead follows (for sufficiently small
instability exponent $\beta$) the exponential law
\begin{equation}
\label{fromopen}
S^{\rm smooth}(E) = {2 \pi \over \beta}
e^{-{\pi \over 2 \beta}|E-\theta|}\,.
\end{equation}
Thus, outside of a small energy window $|E-\theta| \le \beta
|\log \beta|$ surrounding the energy of maximal scarring, most
eigenstates are strongly antiscarred, with mean intensity at the location
of the lead being $S^{\rm smooth} \ll 1$. In the strongly antiscarred
regime where $\Gamma_{an}$ is small, we may ignore the partial
width through the other lead $\Gamma_{bn}$ in Eq.~\ref{defgn}, and so
the conductance $g_n$ is simply proportional to $\Gamma_{an} \sim
S^{\rm smooth}(E_n)$. We then obtain the small-$g$ end of the conductance
distribution:
\begin{equation}
\label{smallg}
P(g) = {2 \beta \over \pi^2 g}\,,
\end{equation}
which holds formally for $\exp(-\pi^2/2\beta) \ll g/s_0 \ll \beta \ll 1$.
The constraint $\exp(-\pi^2/2\beta) \ll g/s_0$ is of no practical
significance; we also note that although Eq.~\ref{smallg} becomes exact
only in the small $\beta$ limit, quantitative agreement is already
obtained for exponents $\beta \approx 0.5$. 
Eq.~\ref{smallg} should be compared with the generic small-$g$ behavior
$P_{\rm generic}(g) = \sqrt{2 /\pi s_0 g}$ predicted by
the Porter--Thomas law (Eq.~\ref{generic}).

\begin{figure}
\psfig{file=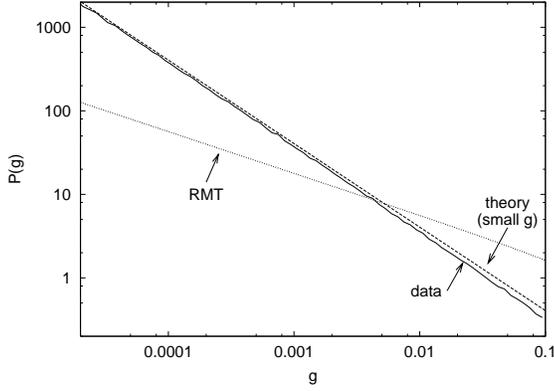,angle=270,width=3in}
\vskip 0.1in
\caption{The small-$g$ part of the peak height distribution $P(g)$ is plotted
on a log-log scale for one lead on an unstable periodic orbit with instability
exponent $\beta=0.2$ (solid curve). The asymptotic prediction of
Eq.~\ref{smallg}, valid for $g \ll \beta \ll 1$ appears as a dashed line of
slope $-1$. For reference, the RMT prediction of Eq.~\ref{generic} appears
as a dotted line of slope $-1/2$. Note the different power-law behavior. Again,
the normalization is set so that the mean partial width through each lead
is given by $s_0=1$. The data would be almost identical for the cases where 
both leads are located on the same periodic orbit with $\beta=0.2$, or where
they are located on two in-phase orbits (not shown).}
\label{fighead}
\end{figure}

The above derivation
applies to an optimal lead ($Q=0$ in Eq.~\ref{aphi}); for a non-optimal
lead the result of Eq.~\ref{smallg} is modified only by
a $Q$-dependent constant, leaving the very distinct
$1/g$ scaling behavior unchanged. The small-$g$ behavior
of Eq.~\ref{smallg} is plotted (for instability exponent $\beta=0.2$)
as a dashed line in Fig.~\ref{fighead}, and differs greatly from
RMT expectations (the latter plotted as a dotted line in the same figure).

The very large fraction of small conductance peaks clearly must
have a significant
effect on the moments of the conductance peak distribution. From
Eq.~\ref{fromopen} we see that only a fraction
$O(\beta |\log\beta|)$ of all energies
have $S^{\rm smooth} \ge 1$ and at these energies we obtain
$g\sim s_\ast \sim s_0$
(see Eq.~\ref{sast1}). At all other  energies we have
$S^{\rm smooth} \ll 1$ and thus $g\sim s_\ast \ll s_0$. All moments
$k \ge 1$
of the conductance distribution then scale as
\begin{equation}
\left\langle \left({g \over s_0}\right )^k \right\rangle \sim \beta
|\log\beta|\,
\end{equation}
for sufficiently small $\beta$.
In particular, the first two moments are given by
\begin{eqnarray}
\langle g \rangle & \approx & 0.16 \beta |\log \beta| s_0 \nonumber \\
\langle g^2 \rangle & \approx & 0.37 \beta |\log \beta| s_0^2 \,.
\label{oneleadmoms}
\end{eqnarray}
The inverse participation ratio (IPR) is a useful dimensionless
measure of the variation in heights which does not
require one to predict theoretically the mean of the distribution:
\begin{equation}
{ \langle g^2 \rangle \over \langle g \rangle ^2} \approx {14.5 \over
\beta |\log \beta|}\,.
\label{oneleadipr}
\end{equation}
This result should be compared with an IPR of $3$ for the generic
Porter--Thomas distribution (see Eq.~\ref{genmoms}). We see a greatly
enhanced fluctuation in peak heights in the case where one of the
leads is located in the periodic orbit. We also note that the numerical
prefactor of $\approx 14.5$ is valid for an optimally placed lead ($Q=0$
in Eq.~\ref{aphi}). In general there is an additional
prefactor which is an easily computable analytic
function of the width of the Gaussian lead mode and of the monodromy
matrix of the periodic orbit, but the important scaling behavior,
i.e.  the increase as $1/\beta |\log\beta|$ of the fluctuations
for small $\beta$, is unchanged. 

\begin{figure}
\psfig{file=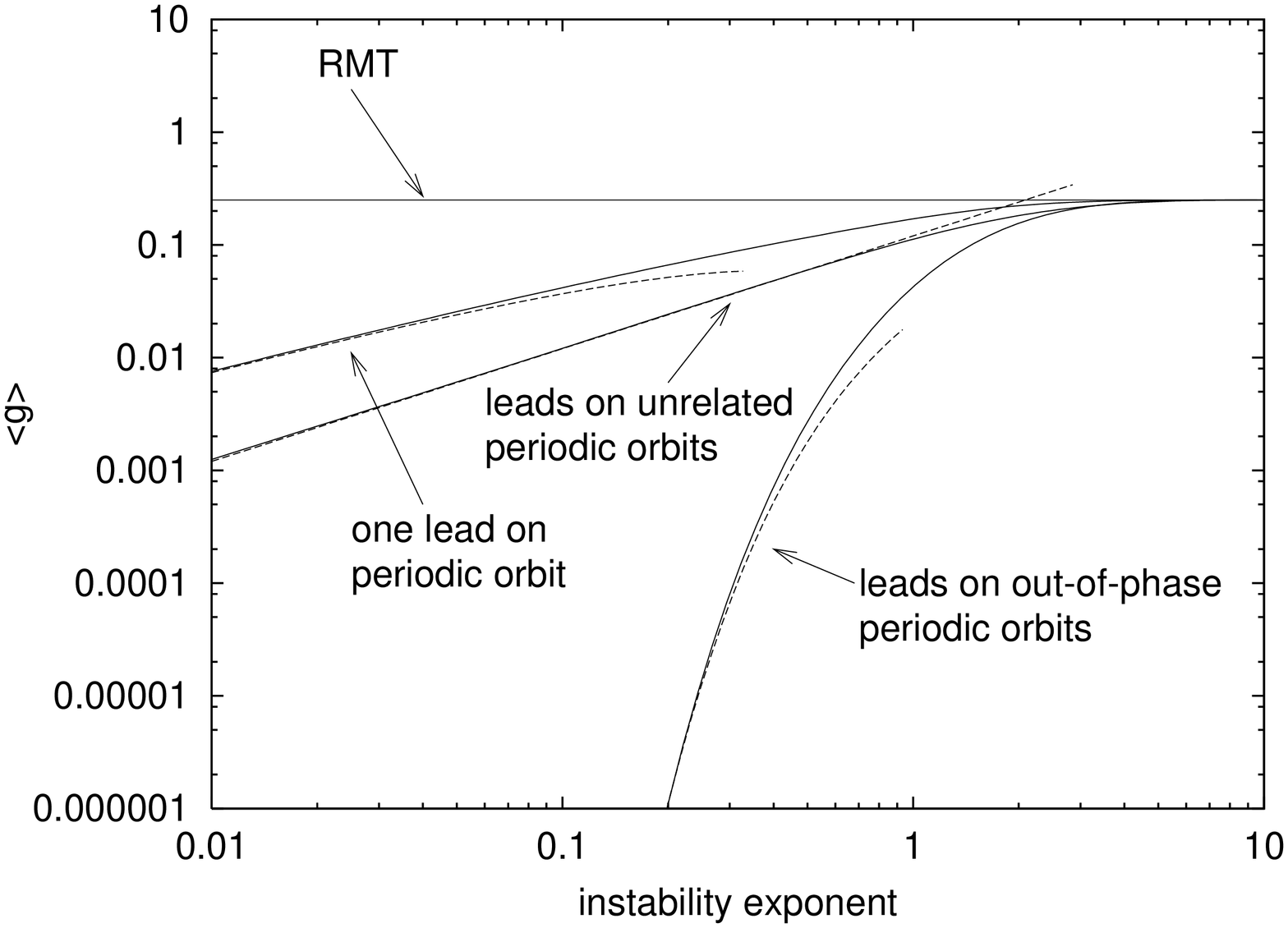,angle=270,width=3in}
\psfig{file=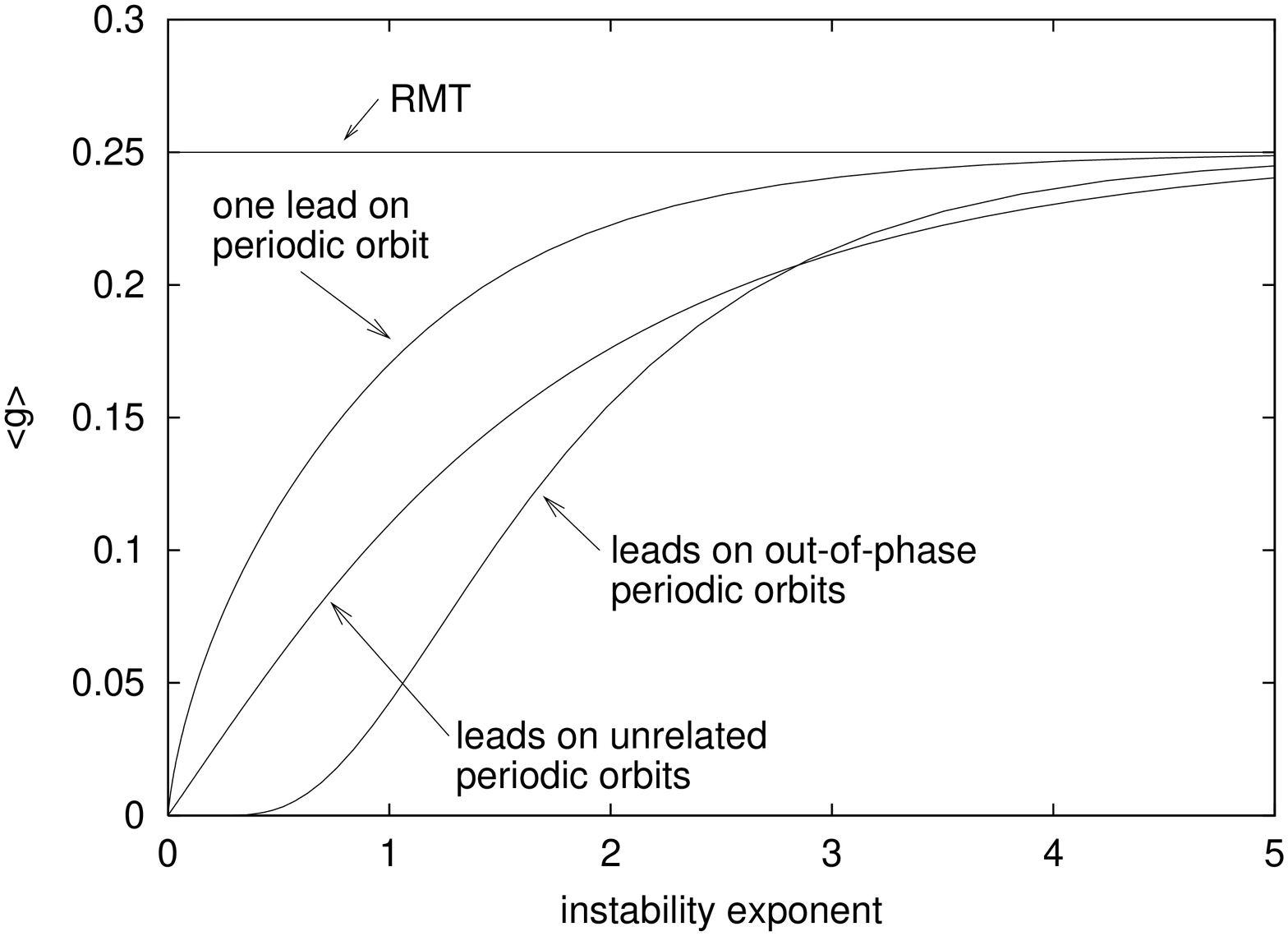,angle=270,width=3in}
\vskip 0.1in
\caption{The mean conductance peak height $\langle g \rangle$ is plotted
as a function of the instability exponent $\beta$. Three cases are shown:
(i) one lead on an orbit of exponent $\beta$, with the other lead placed
generically; (ii) two leads on unrelated orbits each having
exponent $\beta$; and (iii) two leads on out-of phase periodic orbits
with exponent $\beta$. For two leads on the {\it same} periodic orbit,
the mean conductance follows the RMT prediction (with symmetry), and
is independent of the exponent $\beta$.
(Top) Solid curves are exact; dashed curves are
asymptotic forms valid for $\beta \ll 1$, given by Eqs.~\ref{oneleadmoms},
~\ref{meanguncorrel}, and \ref{meangoutofphase}, respectively. In the
dynamics-free limit $\beta \to \infty$, all curves approach the RMT
prediction $0.25$, which is shown on the graph for comparison.
(Bottom) Same data is shown for moderate to large values of the exponent
$\beta$, where the asymptotic forms are not applicable.}
\label{figmean}
\end{figure}

The numerically computed
mean conductance $\langle g \rangle$ for one lead on an orbit
of instability exponent $\beta$ appears in Fig.~\ref{figmean}
as a function
of $\beta$ (solid curve). We observe significant deviations from the RMT value
of $0.25$ (Eq.~\ref{genmoms}, where $s_0$ has been set to unity) for $\beta$
as large as $2.0$, while for larger values of $\beta$ the RMT limit is
approached. The dashed curve in Fig.~\ref{figmean}(top) shows the asymptotic
prediction of Eq.~\ref{oneleadmoms}, which is observed to agree well with the
exact results only for very small $\beta$.

\begin{figure}
\psfig{file=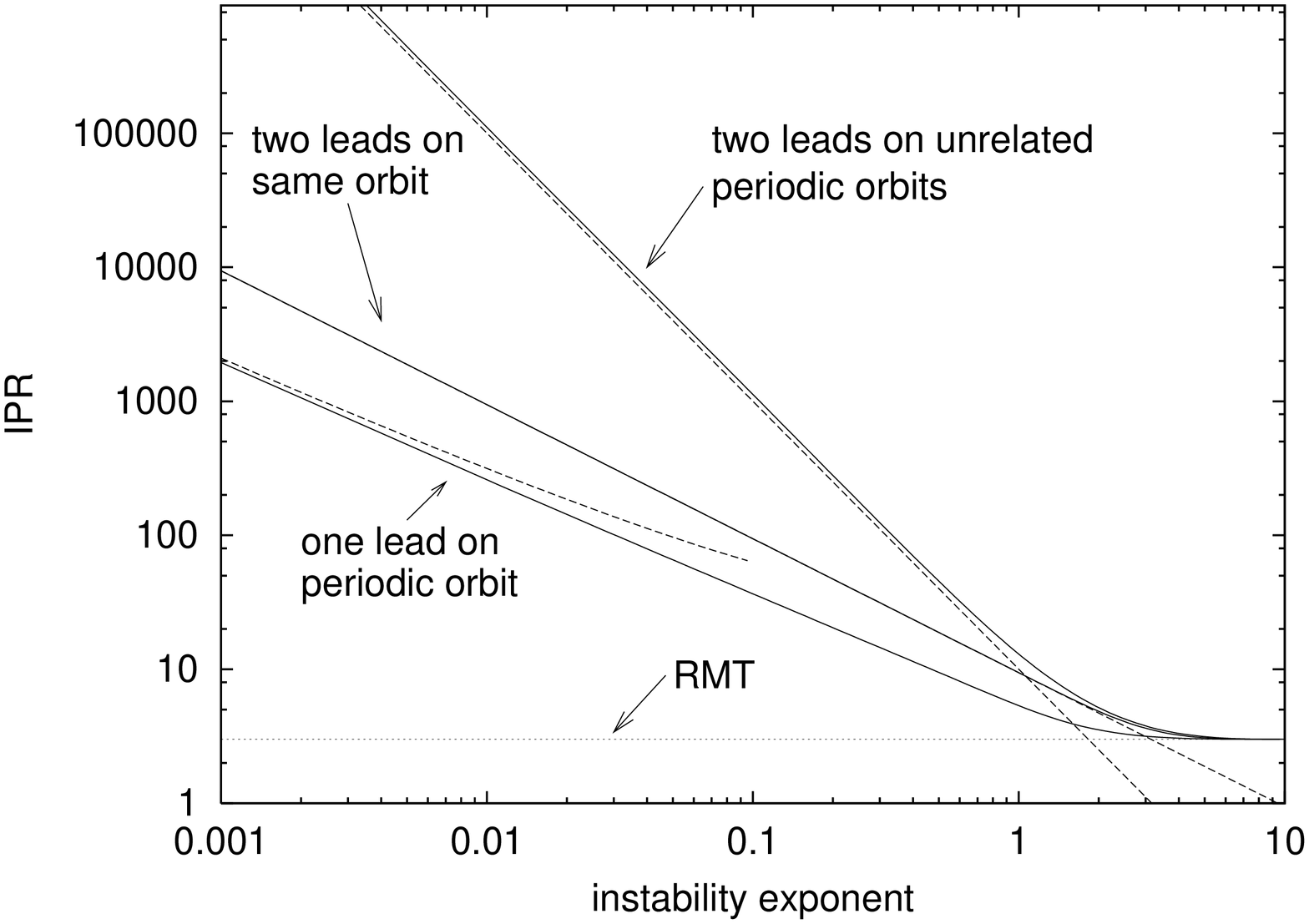,angle=270,width=3in}
\psfig{file=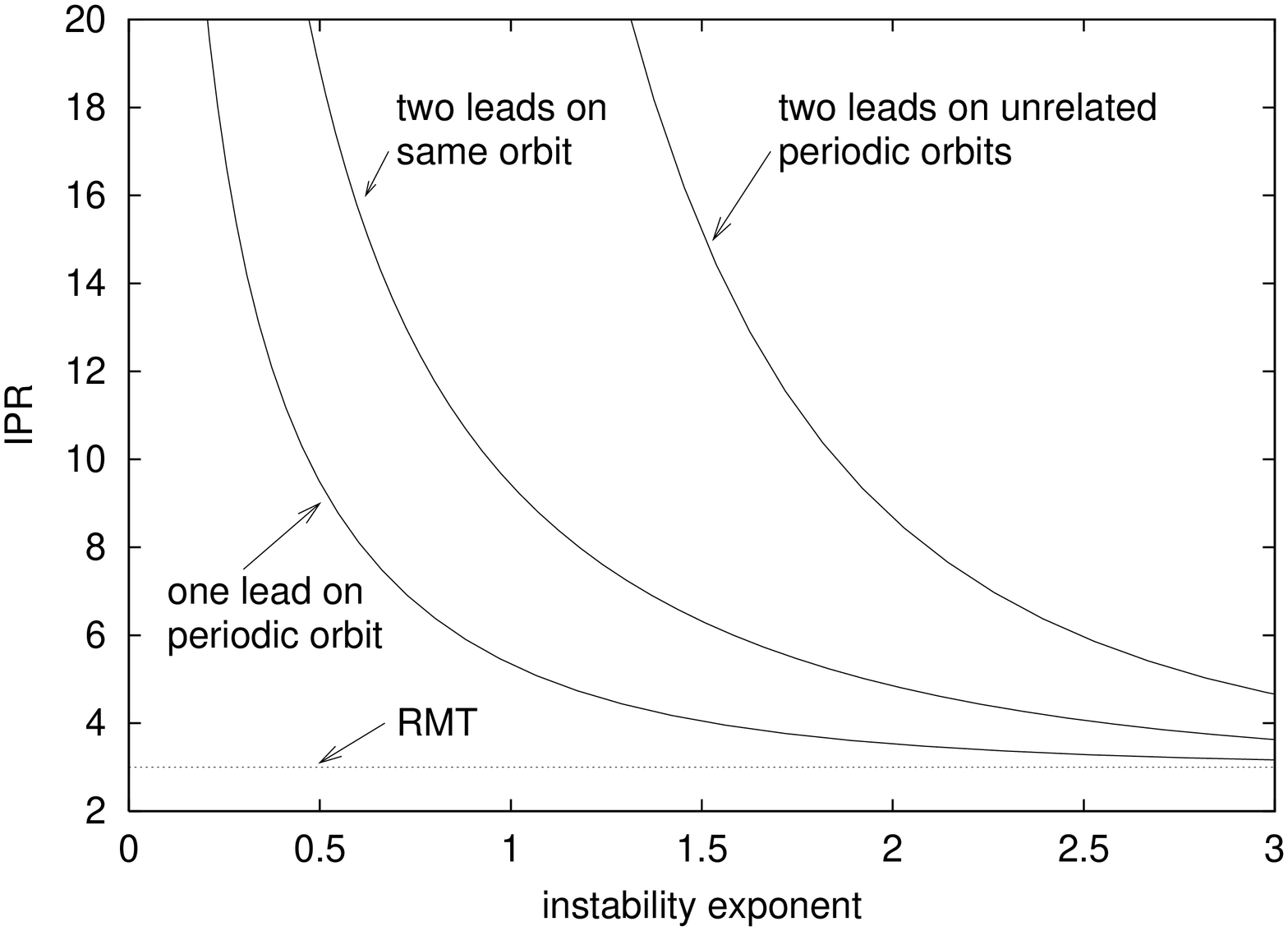,angle=270,width=3in}
\vskip 0.1in
\caption{The IPR (ratio of mean squared conductance
peak height to the square of the mean)
is plotted as a function of the instability exponent $\beta$ of the periodic 
orbit. Three cases are shown: (i) one lead on an orbit with exponent
$\beta$, with the other lead placed generically; (ii) two leads on unrelated
periodic orbits, both with exponent $\beta$; and (iii) two leads on the
same periodic orbit. The RMT prediction of $3$ is shown as a dotted line
for reference; all solid curves converge to the RMT value in the $\beta \gg 1$
regime where the orbit ceases to be important.
(Top) The solid curves are exact numerical results, while
the dashed curves give the asymptotic $\beta \ll 1$ predictions
of Eqs.~\ref{oneleadipr},~\ref{ipruncorrel}, and \ref{iprsameorbit}. (Bottom)
Same data is shown for moderate values of $\beta$, where the asymptotic forms
are not applicable.}
\label{figipr}
\end{figure}

The asymptotic behavior of Eq.~\ref{oneleadipr} appears in
Fig.~\ref{figipr}(top)
as a dashed curve, and can be compared with numerical
data, which is
plotted as a solid line. We observe that the approximations leading
to Eq.~\ref{oneleadipr} do not lead to a quantitatively correct answer
until we reach the very weakly unstable
$\beta \ll 0.1$ regime. On the other hand, strong enhancement of the
IPR compared to the RMT value of $3$ is already clearly visible even near
the moderate exponent $\beta=1$, where the IPR is observed to be
almost twice the RMT prediction. Fig.~\ref{figipr}(bottom) shows
the same calculation,
focusing in on the
IPR behavior for moderate values of $\beta$, where the approximations
leading to Eq.~\ref{oneleadipr} do not apply.

The analysis in this subsection easily generalizes to the
case where the lead located on the periodic orbit also
lies on a symmetry line of the system. As discussed in the previous
subsection, the odd eigenstates do not then produce resonance peaks,
while for the even states we use $s_{a0}=2s_0$, and thus
\begin{equation}
s_\ast(E)=4s_0\left( {1 + 1/\sqrt{2S^{\rm smooth}(E)}} \right)^{-2} \,.
\end{equation}
instead of the expression given by Eq.~\ref{sast1}.

\subsection{Leads on same periodic orbit, or on orbits related by symmetry}
\label{subsamepo}

We now proceed to consider more generally the case discussed by
Narimanov et al.~\cite{narim}, where the two leads are ether located
on the same periodic orbit, or are located on orbits related by
symmetry. First, we notice that if the two leads themselves are 
related by a symmetry, then the two wavefunction intensities and thus
the two partial
widths are identical for each resonance, just as in the discussion
leading to Eq.~\ref{symmetric} in the generic case. Less trivially,
let us consider two leads located on the same orbit but not related
by a symmetry (for example, one may consider the horizontal
bounce in a deformed stadium billiard lacking a left--right symmetry,
and two leads placed at either end of the horizontal bounce orbit).
If the Gaussian packet corresponding to the transverse mode in lead
$b$ were related to the Gaussian packet associated with lead $a$
by time evolution in the closed system, i.e.
\begin{equation}
|\phi_b \rangle = e^{-i\hat Ht/\hbar}|\phi_a \rangle
\end{equation}
for some $t$, then $|\phi_a \rangle$ and $|\phi_b \rangle$ would have
identical local densities of states, $|\langle\Psi_n|\phi_a\rangle|^2=
|\langle \Psi_n|\phi_b \rangle|^2$,
and thus once again $\Gamma_{an}=\Gamma_{bn}$ for each resonance $n$.

More
generally, even though the centers of the Gaussians $|\phi_a \rangle$
and $|\phi_b \rangle$ must be related by time evolution if they lie on the
same orbit, the time-evolved version of $|\phi_a \rangle$ may
have a different aspect ratio or phase space orientation from
$|\phi_b \rangle$. In Ref.~\cite{scarmometer} it was found, however,
that for any two {\it optimally} shaped leads, the two local densities
of states become almost identical in the the limit of
small instability exponent $\beta$. Furthermore the {\it minimum}
possible correlation
between the two partial widths was shown to be $0.94$ even
for $\beta$ as large as $2.3$ (corresponding to a classical
stretching factor $e^\beta=10$), and this correlation
becomes even stronger for smaller
$\beta$, so that for all practical purposes
in the regime where scarring effects are important we
may take $\Gamma_{an} \approx \Gamma_{bn}$ for any two optimally oriented
leads.

In the case where either of the leads $a$ or $b$ are not optimally
shaped so as to be aligned with the stable and unstable manifolds
of the periodic orbit, we have the more general situation
\begin{eqnarray}
\Gamma_{an} &=& C_a (\alpha r_{n} +(1-\alpha)r_{an})S^{\rm smooth}(E_n)
 \nonumber \\
\Gamma_{bn} &=& C_b (\alpha r_{n} +(1-\alpha)r_{bn})S^{\rm smooth}(E_n)\,,
\label{alphaeq}
\end{eqnarray}
where we normally consider equally-coupled leads $C_a=C_b$ as before,
and $r_{n}$, $r_{an}$, and $r_{bn}$ are independent
random (Porter--Thomas) variables. The parameter $0< \alpha< 1$
may be determined in terms of the linearized time evolution
of the two wavepackets around the orbit~\cite{scarmometer}.
In the weakly correlated regime $\alpha \to 0$, the behavior becomes
identical to that of two leads on different orbits but having 
in-phase short-time envelopes $S^{\rm smooth}(E)$.
See Section~\ref{secdifforbit} below for a detailed discussion
of that situation;
for now we restrict ourselves to the strongly correlated case
$\alpha \approx 1$.
We also note that all results valid for two leads on the same
orbit are obviously valid also for leads on two orbits related
by symmetry (e.g. for the two V-shaped orbits of the stadium). 

In the case $\alpha \approx 1$, valid for two nearly optimal leads
located on a weakly unstable orbit, the conductance peak height distribution
reduces simply to the distribution of wavefunction intensities on a short
periodic orbit. This distribution has been studied previously
\cite{wis,antiscar}
in the absence of time reversal symmetry (i.e. in the presence of a magnetic
field); here we need to develop parallel results for the case where
a magnetic field is absent and the wavefunctions are therefore real.

We begin with the moments of the distribution. At any given energy,
the peaks heights $g$ are distributed according to a Porter--Thomas
distribution as in Eq.~\ref{symmetric}, but the mean height ($s_0/2$
in the generic case) must be generalized to $S^{\rm smooth}(E) s_0/2$
(see Eq.~\ref{intfluct}). The moments of $P(g)$ are then given by
\begin{equation}
\langle g^q \rangle = \left\langle \left(S^{\rm smooth}(E)\right)^q
\right\rangle 
\left\langle \left(r_n{s_0\over 2}\right)^q \right\rangle\,,
\end{equation}
where the average in the first factor is over energies $E$ and in
the second factor $r_n$ is distributed as the square of a Gaussian
variable with variance $1$. For generically located (though symmetrically
placed) leads, $S^{\rm smooth}(E)=1$
independent of $E$ and we recover the result of Eq.~\ref{symmetric}.
We notice also that $\langle S^{\rm smooth}(E)\rangle=1$ by normalization,
so the mean conductance peak height is given by
\begin{equation}
\langle g \rangle_{\rm leads \; on \; same \; orbit} = 
{s_0 \over 2}\,,
\label{leadsameorbit}
\end{equation}
when both leads are located on the same orbit,
independent of the stability of the orbit. This is in contrast with our
finding in the previous subsection (Eq.~\ref{oneleadmoms})
that the mean conductance is suppressed
when only one lead is located on a short orbit. Furthermore, 
the mean squared intensity in this case is {\it enhanced} compared to
the generic value:
\begin{eqnarray}
\langle (S^{\rm smooth}(E))^2\rangle &=& \sum_t |A^{\rm short}(t)|^2 \nonumber
\\ &=& \sum_m {1 \over \cosh{\beta m}} \nonumber \\
& \approx & {\pi \over \beta}\,,
\end{eqnarray}
where in the first line we have used the property of Fourier transforms
(recalling Eq.~\ref{ft}), in the second line we have substituted from
Eq.~\ref{aphi} the optimal ($Q=0$) form of the short-time autocorrelation
function, and in the third line we have taken the strong scarring ($\beta
\ll 1$) limit. Then we obtain
\begin{equation}
\langle g^2 \rangle \approx 3 {\pi \over \beta} \langle g \rangle^2\,.
\label{iprsameorbit}
\end{equation}
The validity of this result is confirmed in Fig.~\ref{figipr}(top), where 
the IPR ($\langle g^2 \rangle/\langle g \rangle^2$) is plotted as a
function of the instability exponent $\beta$. The data for two leads on the
same orbit appears there as the middle solid curve, which for 
$\beta \le 2$ agrees very well with the asymptotic prediction of
Eq.~\ref{iprsameorbit} (represented by a dashed line of slope $-1$). The
behavior of the same quantity for larger exponents $\beta$ can be seen
in Fig.~\ref{figipr}(bottom), where we observe that already for
$\beta=1$ the size of peak height fluctuations is three times the RMT
expectation, and that deviations from RMT are noticeable even for $\beta$
as large as $3$.

Thus the IPR
can greatly exceed the RMT value of $3$ and grows as the orbit
on which the leads are placed becomes less unstable. This suggests that
there should be an excess of both very large and very small peaks
as compared with the mean (qualitatively similar to the result
(Eq.~\ref{oneleadipr}) for only one lead on a periodic orbit).
Indeed one can study the tail of the $P(g)$
distribution in a way completely analogous to the calculation
in the previous subsection (and to the similar calculation in~\cite{wis}
for the non-TRS case). From Eq.~\ref{asympt}, we obtain (using
$s_\ast(E)=2s_0 S^{\rm Smooth}(E)$ instead of the corresponding form
Eq.~\ref{sast1} appropriate for only one lead on a periodic orbit) 
\begin{equation}
\label{largegsameorbit}
P(g)=\sqrt{C \over 2\pi^2D} {\beta \over g} e^{-\beta g/Cs_0} \,,
\end{equation}
valid for small $\beta$ and large $g$.
As in Eq.~\ref{oneleadasympt}, we see a much slower exponential decay 
in the tail than that predicted by RMT, but in this case the exponent
is strongly $\beta$-dependent and so the tail becomes ever longer
as $\beta \to 0$. Eq.~\ref{largegsameorbit} (for $\beta=1$)
appears in Fig.~\ref{figtail} as the rightmost dashed curve, and follows
very well the exact
numerical result, which is shown as a solid curve. We see that the asymptotic
prediction of Eq.~\ref{largegsameorbit} is already very good even
for the moderate exponent $\beta=1$, while the RMT prediction
(Eq.~\ref{symmetric}) is completely
wrong in the tail (compare with dotted curve), even after symmetry effects
are included.

This strong enhancement in the number of very large
conductance peaks is balanced
by a corresponding increase in the
number of
very small conductance peaks: the small-$g$ part of the distribution
is given again by Eq.~\ref{smallg}, exactly as in the case
of only one lead on a periodic orbit. The small-$g$ data for two leads on
the same orbit is not shown here, but appears very similar to the case
already plotted in Fig.~\ref{fighead} for only one lead on a short orbit.

The expressions obtained in this subsection need, of course,
to be modified for situations where one of both leads are located
on a symmetry line of the system. The necessary modifications are
completely analogous to those discussed in the previous two subsections,
and we do not go into the details here.

\subsection{Leads on different periodic orbits}
\label{secdifforbit}

We now arrive at perhaps the most interesting situation, where
the two leads are located on distinct (not symmetry-related) periodic
orbits of the classical system. This situation is particularly interesting
because a very rich diversity of behavior may be obtained depending on
the relative phase between the classical actions, in units of $\hbar$,
of the two orbits. For definiteness and for simplicity of presentation,
we will focus on the case where two
orbits have equal stability exponents $\beta$, though the qualitative
results obviously do not depend on this assumption. We also take the periods
of the two orbits to be approximately equal, instead of being related
by some other simple fraction like $1/2$ or $2/3$ (see the end of this
subsection for a discussion of what happens in the case where the two
orbit lengths are not at all simply related). Then the LDOS envelopes
$S^{\rm smooth}(E)$ at the two leads are identical oscillatory functions
of energy, shifted with respect to each other by some phase $\delta \theta$
(see Eqs.~\ref{ft},~\ref{aphi}).

For $\delta \theta=0$, we obtain two in-phase envelopes, and recover the
$\alpha=0$ situation of identical smooth
envelopes with uncorrelated fluctuations
under the envelopes. This scenario was mentioned already in the previous
subsection (Eq.~\ref{alphaeq}), but the discussion there was postponed until
now. We recall that for two uncorrelated partial widths with the same mean,
the distribution of conductances is Porter--Thomas (Eq.~\ref{generic}),
just as in the perfectly correlated case (Eq.~\ref{symmetric}), but the
mean conductance in the uncorrelated case is only half as large as 
in the correlated case. This argument applies of course independently
at each energy, so even after averaging over energy we obtain the
same distribution of conductances as in the previous subsection, but
with the mean value shifted by a factor of $2$. To summarize,
the mean is given by 
\begin{equation}
\langle g \rangle_{\rm leads \; on \; in-phase \; orbits} = 
{s_0 \over 4}
\end{equation}
(cf. Eq.~\ref{leadsameorbit}), while the IPR and the small-$g$ part
of the distribution (Eqs.~\ref{iprsameorbit},~\ref{smallg}),
respectively, remain unchanged
from the previously considered case. The tail of the conductance peak
distribution becomes
\begin{equation}
\label{largeginphaseorbit}
P(g)=\sqrt{C \over 2\pi^2D} {\beta \over g} e^{-2\beta g/Cs_0} \,.
\end{equation}

The result of Eq.~\ref{largeginphaseorbit} is plotted for $\beta=1$
as the middle dashed curve in Fig.~\ref{figtail}, and is observed to agree
well with the numerical calculation, given by the corresponding solid curve.
We again note that the asymptotic form of Eq.~\ref{largeginphaseorbit}
works quantitatively even though
$\beta$ in our case is not small; furthermore the tail of the peak height
distribution $P(g)$ again differs greatly from the RMT prediction (leftmost
dotted curve in the same figure).

We now consider two envelopes $S^{\rm smooth}$ that are not in phase with
each other. From Eq.~\ref{fromopen} we know that each envelope falls off
exponentially away from its peak, and we also know from Eq.~\ref{defgn}
that when the two local densities of states are unequal, the conductance
is always dominated by the more weakly-coupled lead (smaller $S^{\rm smooth}$).
The maximum conductance then occurs at energies where
$S_a^{\rm smooth} \approx S_b^{\rm smooth}$. This maximum occurs at
$E=(\phi_a+\phi_b)/2 \; {\rm mod} \; 2\pi=
\phi_a+\delta \phi/2 \; {\rm mod} \; 2\pi$, at which energy
$S_a^{\rm smooth} =  S_b^{\rm smooth} =(2\pi/\beta)\exp(-\pi \delta \phi
/4 \beta)$. The {\it maximum} possible conductance peak height is
therefore very strongly suppressed compared with the RMT prediction:
\begin{equation}
g^{\rm max}_{\rm out \; of \; phase \; orbits} \sim 
{s_0 \over \beta} e^{-\pi \delta \phi /4 \beta} \,,
\end{equation}
where the out-of-phase parameter $\delta \phi$ should be taken between
$0$ and $\pi$. We also note that the fraction of energy space at which
conductances this large are attained scales as $\beta$ because of the
exponential falloff in Eq.~\ref{fromopen}; thus  the {\it mean} conductance
peak height scales as
\begin{equation}
\langle g\rangle_{\rm out \; of \; phase \; orbits} \sim  
s_0 e^{-\pi \delta \phi /4 \beta} \,,
\label{meangoutofphase}
\end{equation}
suppressed by an exponential factor from the generic expectation
of Eq.~\ref{genmoms}. The most dramatic effect arises of course
when the two envelopes are out of phase by exactly $\pi$ in the energy
region of interest. There we have
\begin{equation}
g^{\rm max} \sim {s_0 \over \beta}  e^{-\pi^2/4\beta} \;\;\;\;
\langle g\rangle \sim s_0 e^{-\pi^2/4\beta}\,.
\label{meanoutofphase}
\end{equation}
An exact numerical calculation for $\langle g \rangle$ is plotted as the
lowest solid curve in Fig.~\ref{figmean}(top), and is observed to approach
the asymptotic prediction of Eq.~\ref{meanoutofphase} for small $\beta$,
while for large $\beta$ it of course tends to the RMT prediction
of $0.25$. Again, Fig.~\ref{figmean}(bottom) shows in more detail the behavior
for moderate-to-large values of $\beta$.

This situation described by Eq.~\ref{meanoutofphase}
may be obtained in one of the following two ways. One
possibility is to take two orbits of the same period but different Maslov
phases, resulting in out-of phase behavior at all energies. Alternatively
one may consider two orbits of only approximately equal period; then
the two oscillating envelopes move in and out of phase with each other
as one sweeps through energy by adding electrons to the dot.

In the scenario of leads placed on
two orbits with irrationally-related periods, we can also imagine
collecting statistics over an energy window wide enough to include both
in-phase and out-of-phase behavior. We see in this case that the peak
height will always be exponentially suppressed except at those energies
where {\it both} LDOS envelopes $S^{\rm smooth}_a(E)$
and $S^{\rm smooth}_b(E)$
are near their respective maxima, i.e. where the wavefunction is scarred
simultaneously at each of the two leads. Since in each envelope the bumps
corresponding to scarred wavefunctions have width $O(\beta)$ compared with
the spacing between the bumps, and since the bumps in 
$S^{\rm smooth}_a(E)$ and $S^{\rm smooth}_b(E)$ are now assumed to be
uncorrelated, we find that a fraction $O(\beta^2)$ of all wavefunctions
give rise to substantial peaks, of height $O(\beta^{-1} s_0)$. Then the
mean conductance peak height scales for small $\beta$ as
\begin{equation}
\langle g\rangle_{\rm uncorrelated \; orbits} \sim  
\beta s_0 \,,
\label{meanguncorrel}
\end{equation}
and the higher moments of the distribution behave similarly:
\begin{equation}
\langle g^k\rangle_{\rm uncorrelated \; orbits} \sim      
\beta^{2-k} s_0^k \,;
\end{equation}
\begin{equation}
\langle g^{\rm max}\rangle_{\rm uncorrelated \; orbits} \sim    
\beta^{-1} s_0 \,;
\end{equation}
Again, we may measure the fluctuation in conductance peak heights
by taking the
ratio of the mean squared peak height to the square of the mean (IPR).
This shows very strong deviations from RMT expectations:
\begin{equation}
\left({\langle g^2\rangle \over \langle g\rangle^2} \right)_{\rm
uncorrelated \; orbits} \sim \beta^{-2}\,.
\label{ipruncorrel}
\end{equation}
Because both leads must be scarred to obtain appreciable conductance,
the fluctuation in conductance peak heights is much stronger here than
in the case of only one lead on a periodic orbit (Eq.~\ref{oneleadipr}) or even
in the case of two leads on the same orbit (Eq.~\ref{iprsameorbit}).

The mean conductance peak height $\langle g \rangle$ for leads on two
unrelated periodic orbits of instability exponent $\beta$ is plotted 
as a function of $\beta$ as the labeled solid curve in
Fig.~\ref{figmean}(top), where the $\beta \ll 1$ asymptotic prediction
of Eq.~\ref{meanguncorrel}
appears as a dashed line. The behavior for larger $\beta$
again can be seen in Fig.~\ref{figmean}(bottom). Similarly, the IPR
for this case can be found plotted numerically by the correspondingly
labeled solid curves in Fig.~\ref{figipr}(top and bottom); again
for small $\beta$ the result agrees with the asymptotic prediction
of Eq.~\ref{ipruncorrel} (Fig.~\ref{figipr}(top); dashed line).

\subsection{Leads with unequal coupling}
\label{secunequal}

We have been focusing throughout on the simple and
experimentally motivated case
of two equal-sized leads; however all of the discussion and calculations
in this section
generalize in a very straightforward way to a scenario with unequally
coupled leads.
Qualitatively, the main observation that should be added to the previous
discussion is that the lead which is more weakly coupled naturally
has a stronger effect on the conductance peak heights and their distribution,
and the location of the more weakly coupled lead relative to classical
structures is most important in understanding the conductance behavior.

A particularly interesting case to consider is where one of the leads,
say $b$, is coupled to the dot much more strongly than the other lead,
so $C_a \ll C_b$ in Eq.~\ref{gammapsi}. Then Eq.~\ref{defgn} reduces
to $g_n \approx \Gamma_{an}$, and the conductance depends only
on the intensity of the wavefunction near lead $a$. Since $s_b \gg s_a$,
Eq.~\ref{sast} becomes $s_\ast=4 s_a$, and we obtain behavior identical
to that observed previously
for two symmetrically placed leads (Section~\ref{subsamepo}), except
that $s_0$ must be replaced everywhere by $2 s_a$. Thus, for example,
we obtain 
\begin{equation}
\langle g \rangle = s_a
\end{equation}
(independent of the location of lead $a$), while the IPR
is given by Eq.~\ref{iprsameorbit} if lead $a$ is located on an
orbit of instability exponent $\beta \ll 1$. The tail of the peak height
distribution becomes
\begin{equation}
P(g)=\sqrt{C \over 2\pi^2D} {\beta \over g} e^{-\beta g/2Cs_a}\,.
\end{equation}
Of course, for these results to be valid the coupling to lead $a$ must
be weaker
than the coupling to $b$ even near energies where lead $a$ is scarred;
assuming $b$ is placed generically, this means $C_a \ll \beta C_b$. If this
condition is satisfied, the unequal-leads experiment may present an
practical alternative to trying to ensure that both leads are centered
on a periodic orbit.

\section{Numerical tests}
\label{secnumtests}

We now proceed to look at the implications of the results of the previous
section for a specific chaotic system, namely the Bunimovich stadium. This
consists of two semicircular endcaps of radius $1$, attached to
either side of a 
rectangle of dimension $2 \times 2 \gamma$. The system reduces to a circle
for $\gamma=0$; maximum chaos is attained at $\gamma=1$. We will focus 
mostly on the $\gamma=1$ special case, though scar effects are stronger
at smaller values of $\gamma$ (as the periodic orbit instability
exponents $\beta$ become
smaller).

The (Dirichlet) wavefunctions of a billiard system such as the stadium
can be obtained by the plane wave method, where at each $k_F$ one constructs
the linear combination of plane waves that minimizes the integral
of $|\psi|^2$ along the boundary, and then picks out those values
of $k_F$ where this integral dips down towards
zero. Of course, the integral never becomes strictly zero because of the finite
number of plane waves used at any given $k_F$; however, it can easily
be checked that identifying all the sharp minima in the boundary integral
with eigenvalues produces the right density of states, as predicted by the Weyl
law and higher-order corrections. For each wavefunction obtained using this
method, we can compute the partial widths through each of the two leads
via Eq.~\ref{gammapsi}, by simply integrating the normal derivative of
the wavefunction at the boundary, multiplied by the (Gaussian) lowest
transverse mode in the lead (Eq.~\ref{leadmode}). The conductance peak height
associated with that resonance is then easily obtained using Eq.~\ref{defgn}.

We begin by considering two leads symmetrically placed on the two semicircular
endcaps. As we have seen above
in Eq.~\ref{leadsameorbit}, the mean conductance
peak height should be independent of classical structures near the
lead location, and this behavior is indeed observed when averaging over
the energy range $300 <k_F <350$. However, the fluctuation in peak heights,
as measured for example by the mean squared height, is expected to
depend strongly on whether the leads are located on a short periodic orbit
that is normally incident at the location of the
lead (Eq.~\ref{iprsameorbit}).

\begin{figure}
\psfig{file=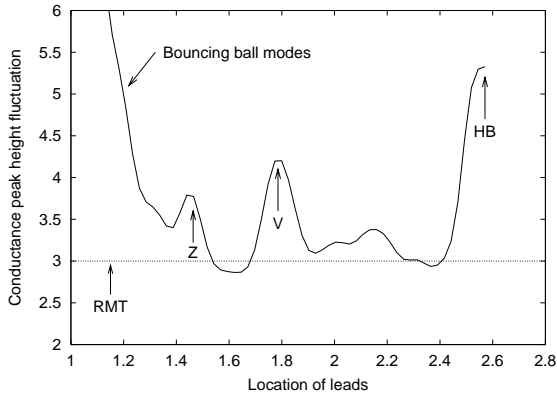,angle=270,width=3in}
\vskip 0.1in
\caption{The IPR of the conductance peak heights (mean squared peak
height divided
by the square of the mean) is plotted as a function of the distance of each
lead from the center of the straight segment in a stadium billiard. The two 
leads
are placed symmetrically on the two semicircular caps, and have width 
$\sigma=1.2/\sqrt k_F$, which corresponds to $\approx 3.5 \lambda_F$ in the
energy range considered ($300 < k_F <350$). The amount of fluctuation
is significantly enhanced when the leads are located on unstable
periodic orbits, as compared with the RMT prediction of $3$. The locations
of the three shortest
periodic orbits having normal incidence on the semicircular endcaps (horizontal
bounce, V-shaped orbit, and Z-shaped orbit) are marked on the plot. On the
horizontal axis, $\gamma=1.0$
corresponds to the end of the straight segment, and 
$\gamma+\pi/2\approx 2.57$ corresponds to the middle of the circular endcap.
}
\label{figiprfcn}
\end{figure}

This qualitative expectation is confirmed in Fig.~\ref{figiprfcn}, where
we clearly see that the mean squared peak height is strongly enhanced over the
RMT value whenever the leads (of width $\sigma=1.2/\sqrt k_F \approx
3.4 \lambda_F$) are located on a short periodic orbit. 
Furthermore, the three shortest (and least unstable)
orbits in the stadium having normal incidence 
on the endcaps are the horizontal
bounce (HB),
the V-shaped orbit, and the Z-shaped orbit, in that order, and these
are seen to correspond precisely to the three most pronounced peaks in the
plot. We note that the periodic orbit length (as well as the
instability exponent $\beta$) 
must be measured by identifying the four quadrants of the stadium billiard,
due to symmetry: thus, for example, the length of the horizontal bounce
orbit is $2+2\gamma=4$. Looking at the length of the orbit (and instability
exponent) in the full billiard, we would be led to underestimate the true
importance of scarring effects.

Notice that the leads are always located on the upper half of the stadium,
equidistant from the center point of the upper straight segment.
Thus, for both the horizontal bounce and the V-shaped orbit, the two leads 
happen to be on the same orbit, while in the case of the Z-shaped orbit, the
leads would be placed on distinct orbits that were related by symmetry.
As noted above, this difference has no effect on the conductance behavior.

We also see a strong increase in peak height fluctuations as the leads approach
the edge of the
straight segment. This is not surprising since the quantum behavior there
is strongly influenced by bouncing-ball and near-bouncing-ball
modes~\cite{bb}. For a lead located {\it on} the straight segment of the
boundary, the IPR will tend to infinity in the high-energy limit. In other
words, a smaller and smaller
fraction of all modes will give rise to appreciable
conductance in this limit.

To understand the conductance behavior at a quantitative level, we focus in
on the most pronounced peak in Fig.~\ref{figiprfcn}, corresponding
to leads located on the horizontal-bounce orbit, i.e. centered opposite to
one another on the
two semicircular endcaps. For $\gamma=1$, the system previously considered
in \cite{narim}, the mean squared peak height is observed to be
enhanced by a factor of $1.78$ over the RMT prediction, as compared with
an expected enhancement of $1.81$ coming from Eq.~\ref{iprsameorbit},
where we have used the value $\beta=1.76$ appropriate to this orbit.
[We note that for the value of the lead width $\sigma$ given above, the
lead shape is close enough to being optimal that we may ignore corrections
associated with non-zero parameter $Q$ in Eq.~\ref{aphi}, discussed above.]
Such close agreement is surely accidental, as one expects deviations
due to the finite number of conductance peaks being sampled, as well
as systematic finite-energy errors associated with corrections to
the semiclassical approximation and also with the presence of non-random
bouncing
ball and whispering gallery modes. Indeed, looking at leads
centered on the horizontal bounce
orbit in the $\gamma=1/2$ stadium, and also at leads centered on the 
V-shaped orbit in the original $\gamma=1$ stadium, we find in each case that
the observed mean squared peak height is about $10\%$ lower than the predicted
value (but still much higher than the RMT expectation). See Table I for details.
In all three cases, the measured value of $\langle g^2 \rangle/ 
\langle g \rangle ^2$ is easily distinguishable from the IPR value of $3$.

\begin{table}
\begin{tabular}{|c|ccc|ccc|}
lead locations & & $\langle g \rangle$ & & & $\langle g^2 \rangle/\langle g
\rangle^2$ & \\ \hline
& actual & theory & RMT & actual & theory & RMT \\ \hline \hline
both on HB & 0.50 & 0.50 & 0.50 & 5.33 & 5.43 & 3.00 \\ \hline
both on HB & 0.50 & 0.50 & 0.50 & 6.44 & 7.17 & 3.00 \\ 
($\gamma=0.5$)  & & & & & & \\ \hline
both on V & 0.50 & 0.50 & 0.50 & 4.23 & 4.80 & 3.00 \\ \hline
HB and & 0.21 & 0.22 & 0.25 & 3.70 & 3.73 & 3.00 \\
generic &&&&&& \\ \hline
HB and V & 0.20 & 0.19 & 0.25 & 4.11 & 4.50 & 3.00 \\ \hline
generic & 0.50 & 0.50 & 0.50 & 3.02 & 3.00 & 3.00 \\
(symmetric) &&&&&&\\ 
\end{tabular}
\vskip 0.1in
\caption{Mean and mean squared conductance peak heights are shown for 
several lead locations in the stadium billiard, and are compared with
scar theory as well as with RMT. Unless stated otherwise, $\gamma=1$
is used for the shape of the billiard (see text above). We consider two cases
where both leads are located on the same (horizontal bounce) orbit, and one
where they are located on symmetry-related (V-shaped) orbits. In addition,
scenarios are shown where one lead is located on the horizontal bounce,
while the other is placed either on the V-shaped orbit or at a generic
location ($q=2.3$ in the coordinates of Fig.~\ref{figiprfcn}). For reference,
we also show a calculation where the two leads are symmetrically placed at 
a generic location; there we see that good agreement with RMT is obtained.
}
\label{stadtable}
\end{table}

\begin{figure}
\psfig{file=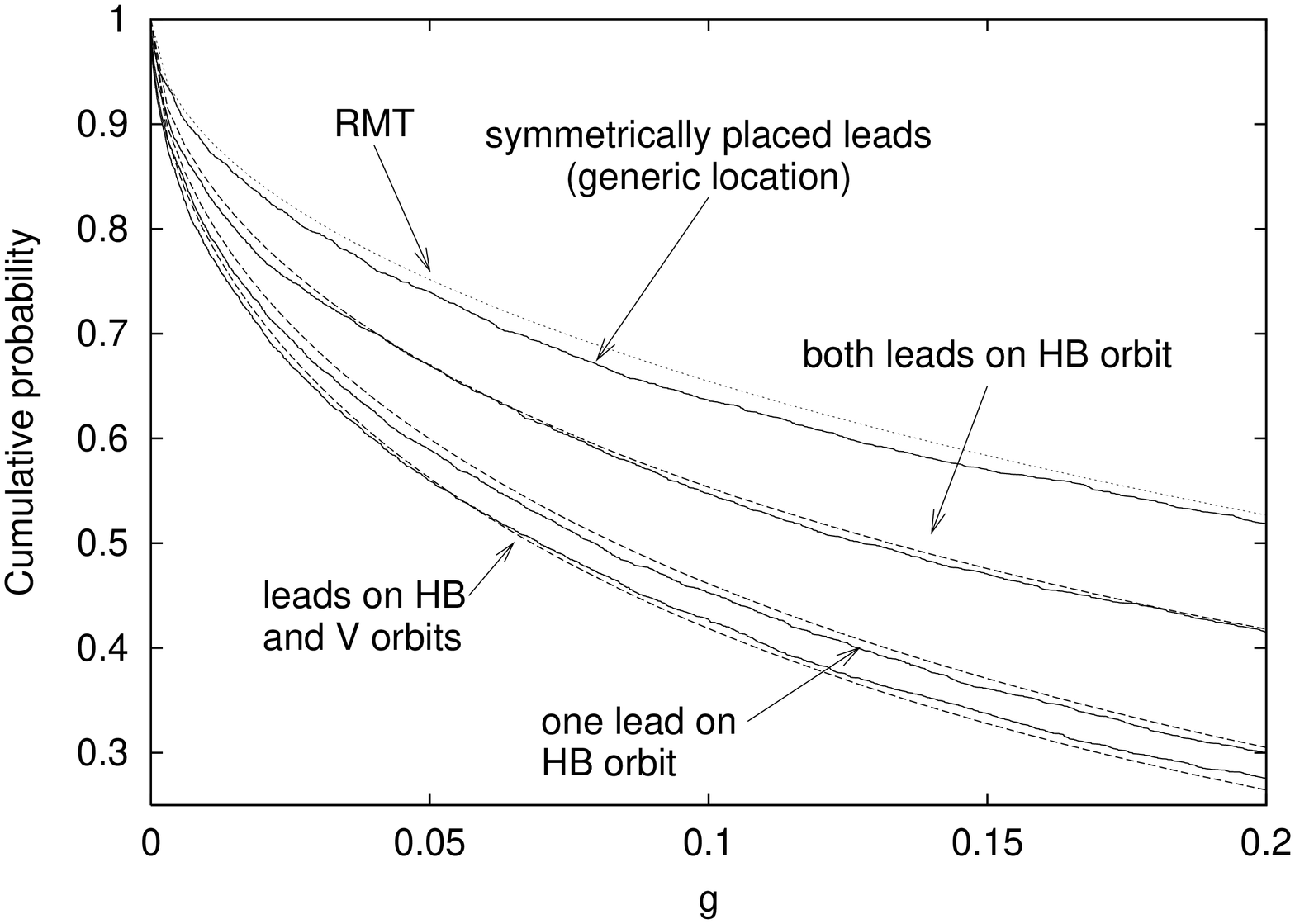,angle=270,width=3in}
\psfig{file=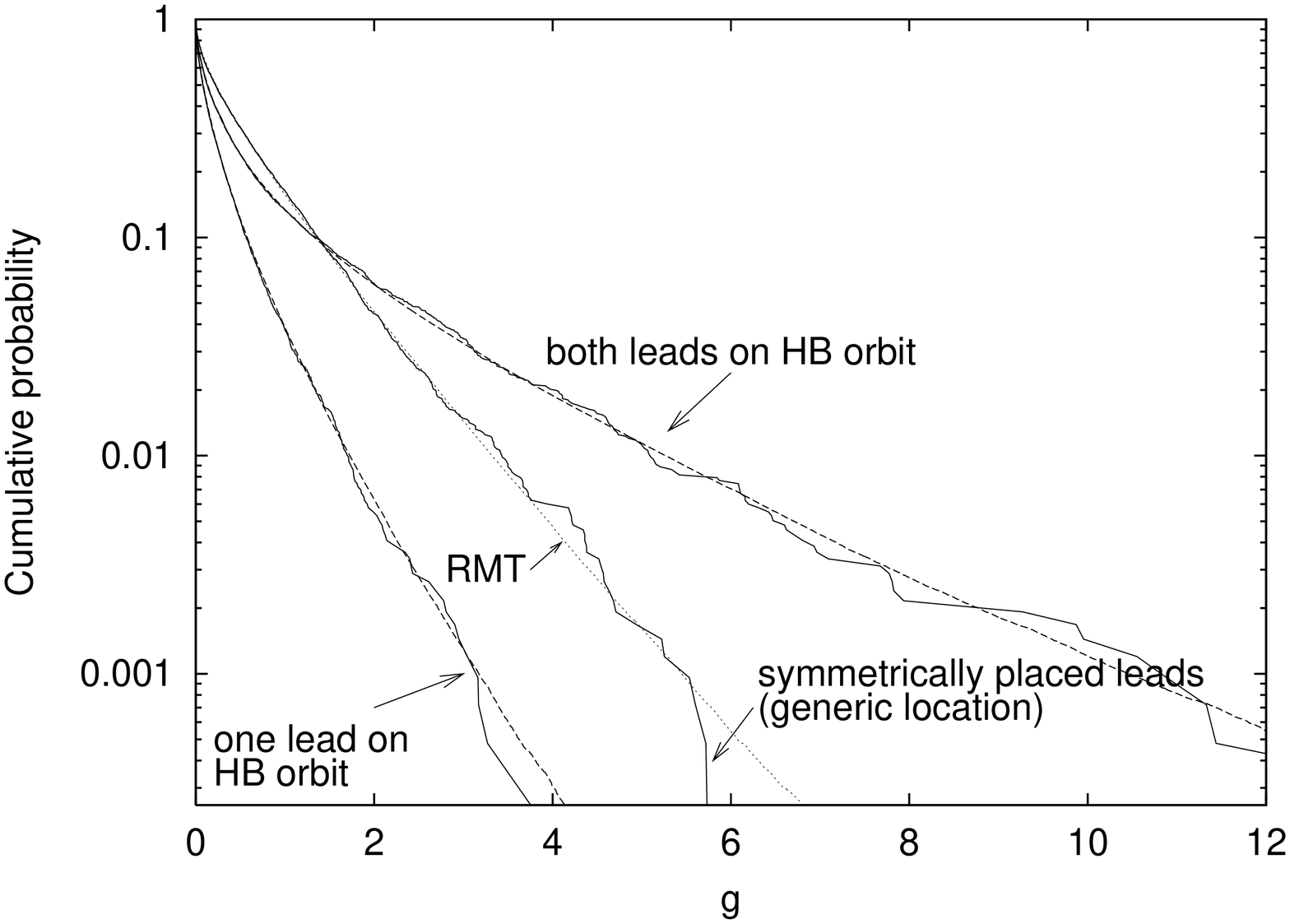,angle=270,width=3in}
\caption{The cumulative probability $\int_g^\infty dg' P(g')$
of having conductance peak height greater than $g$ in a stadium
billiard with two leads is plotted for several lead locations,
illustrating the
various scenarios considered in the previous section: (i) both leads located
on the horizontal bounce orbit, (ii) one lead on the HB orbit and the other
at a generic location, (iii) one lead on the HB orbit and the other on the
V-shaped orbit (top figure only), and (iv) the two leads
symmetrically placed at a generic location. The solid curves represent stadium
data, while the dashed curves represent scar theory predictions
(Eq.~\ref{ponelead}, etc.); for the case of generically placed leads,
the scar
theory prediction coincides with RMT (shown as a dotted curve). The axes
are scaled so as to focus on the tail of the distribution in the bottom figure
and the head in the top figure.
}
\label{figstad}
\end{figure}

The first two moments of $P(g)$, as shown in 
Table~\ref{stadtable}, provide useful and concise information
about the distribution of conductance peak heights $g$, and can be easily
used to distinguish generic lead locations from the case of leads located
on a whost periodic orbit. More complete information is of course contained in
the full distribution, as shown in Fig.~\ref{figstad}. [The distributions
there are shown as cumulative probabilities $\int_g^\infty dg' P(g')$, so
as to reduce the effect of statistical noise.] We see,
for the case of both leads located on the HB orbit, a long tail
(Fig.~\ref{figstad}(bottom), top solid curve) in
good quantitative agreement with the prediction of scar theory (dashed curve). 
We also see in Fig.~\ref{figstad}(top, solid curve, second from top)
the increased number of small
conductance peak heights as compared with RMT, and again in very good
agreement with scar theory predictions (accompanying dashed curve).

We recall also (from Section~\ref{secunequal})
that, up to an overall scaling factor
of $2$, exactly the same distribution
would be obtained for only one lead located
on a short orbit of the stadium, with the other, much more strongly coupled,
lead located at a generic location.

We may also consider attaching two leads to the stadium so that only one
of them is located on a short orbit (compare Section~\ref{seconelead}).
We see from Table~\ref{stadtable} that in this case the mean conductance is
reduced as compared with RMT, while the ratio of the mean square to the square
of the mean is enhanced, in agreement with the predictions. The full
probability distribution for this case also appears in Fig.~\ref{figstad}
(appropriately labeled solid curves, top and bottom), where we see a very
noticeably shortened tail as well as an
excess of small intensities, as compared
with the previously considered case (namely, both leads on the HB orbit). The
distribution, in the head, body, and tail, is
in good agreement with the theoretical prediction of
Eq.~\ref{ponelead}. 

Finally, the next-to-last line in Table~\ref{stadtable} shows the conductance
behavior for the case of two leads found on unrelated short orbits,
in this case the horizontal bounce and the V-shaped orbit. Both the suppression
of the mean conductance and the enhancement of conductance fluctuations
(as compared with RMT) are observed, as predicted in
Section~\ref{secdifforbit}. However, the size of the effect is $\approx 20\%$
smaller than predicted by the theory. This difference
may easily be ascribed to finite-energy effects considered above, in conjuction
with statistical fluctuations. The head of the distribution is shown
in Fig.~\ref{figstad}(top, lowest solid curve), and follows the scar theory
prediction, which again appears as a dashed curve. The tail of the distribution
would overlap the one discussed in the immediately preceding paragraph (one
lead on HB, with the other lead generic), and is not shown.

Finally, the last line in Table~\ref{stadtable} shows the moments of the
conductance peak height distribution for generically placed leads, which are
seen to agree very well with RMT statistics. The distribution for this case
is also plotted in Fig.~\ref{figstad}(top and bottom), and as expected
it follows the
RMT prediction.

\section{Conclusion}
\label{secconc}

In this paper we have studied
the statistical properties of
transmission through a weakly open chaotic cavity. We have seen
that RMT does a very good job in describing these properties 
when the leads
are located far from classically important structures. However,
the random-wave hypothesis
needs to be modified for a quantitatively accurate description of the
transmission process when one or both leads are located near a classically
important structure, such as an unstable periodic orbit. The unstable
periodic orbits have a strong effect on the distribution of conductance peak
heights, even in the limit where the dwell time inside the dot becomes very
large compared to the Lyapunov scale on which classical decay away from the
unstable orbits takes place. This implies that short-time classical dynamics
leaves its imprint on quantum behavior at arbitrarily long time scales, even
though at these scales the classical dynamics loses all memory of its
short time behavior. We emphasize that the effect considered in this paper
is unrelated to direct processes, and arises purely from quantum mechanical
interference.

We have also seen that scar-like effects on conductance behavior in
a quantum dot are robust, and that statistical predictions about the 
distribution of peak heights can be made even though we do not have enough
information to compute the individual resonances of the system, either
quantum mechanically or semiclassically. This means in particular that
changes in the mean field which arise as electrons are added to the dot
may not significantly effect the statistical properties of the conductance,
even though they completely destroy all predictive power for individual
resonant wavefunctions. Similarly, electron-electron interactions are expected
to make the single-particle picture irrelevant for the purpose of predicting
individual resonances or the spacing distribution, but should not affect
the statistical properties of the conductance as long as the interaction
mean free path is larger than the ballistic path associated with one crossing
of the dot.

The mean conductance, higher moments, and the full distribution of peak
heights may all be easily computed using scar theory ideas, for the various
scenarios considered in this paper. These include a single lead located on
a short orbit, two leads located on distinct orbits, or two leads located on
the same orbit. The last scenario was seen to be mathematically equivalent
both to two leads located on orbits related by symmetry, and also to a
situation of unequally-coupled leads, where only the weakly coupled
lead needs to be located on a short orbit. All these scenarios may be relevant
to different experimental setups.

Experimentally, it is of course much easier to change the classical dynamics
inside the dot (either by adjusting the classical geometry or by turning on
a strong magnetic field) than it is to sweep through different locations of the
lead. Thus, one may consider a situation where the lead locations are fixed
but the classical dynamics is changed by adjusting either a voltage or
a magnetic field strength. Changes in the peak height fluctuations are then
predicted when the adjustable 
parameter passes through values which cause a periodic
orbit to hit one or both of the leads.

In this paper we have not considered peak height correlations~\cite{narim}
or the effect
of finite temperature on conductance peak heights; both of these are
important effects which have been addressed previously by other authors.
We have emphasized the crucial importance of including dynamical information
for a quantitative understanding of conductance peak heights
in a quantum dot.
We also note that similar dynamical effects should be present in other
physical situations where tunneling in or out of a chaotic
well is important, and dynamical structures inside the well (such as unstable
periodic orbits) will influence the distribution of coupling strengths between
the well and the outside world. Work along these lines involving tunneling
into a smooth potential well is currently under development.

\section{Acknowledgments}

This research was supported by the DOE under Grant
DE-FG-06-90ER40561.


\begin{references}

\bibitem{quandot} M. A. Kastner, {\it Rev. Mod. Phys.} {\bf 64},
849 (1992).

\bibitem{coulblock} H. Grabert and M. H. Devoret,
{\it Single Charge Tunneling: Coulomb Blockade Phenomena
in Nanostructures} (Plenum, New York, 1992).

\bibitem{narim} E. E. Narimanov, N. R. Cerruti, H. U. Baranger,
and S. Tomsovic, {\it Phys. Rev. Lett.} {\bf 83}, 2640 (1999).

\bibitem{matrelem} T. S. Monteiro, D. Delande, A. J. Fisher, 
and G. S. Boebinger, {\it Phys. Rev.} {\bf B 56}, 3913 (1996);
E. E. Narimanov, A. D. Stone, and G. S. Boebinger, {\it Phys. Rev. Lett.}
{\bf 80}, 4024 (1998); E. B. Bogomolny and D. C. Rouben,
{\it Europhys. Lett.} {\bf 29}, 7 (1995).

\bibitem{rmtcond} R. A. Jalabert, A. D. Stone, and Y. Alhassid,
{\it Phys. Rev. Lett.} {\bf 68}, 3468 (1992); V. N. Prigodin,
K. B. Efetov, and S. Iida, {\it Phys. Rev. Lett.} {\bf 71},
1230 (1993); E. R. Mucciolo, V. N. Prigodin, and B. L. Altshuler,
{\it Phys. Rev.} {\bf B 51}, 1714 (1995); Y. Alhassid and C. H. Lewenkopf,
{\it Phys. Rev. Lett.} {\bf 75}, 3922 (1995).

\bibitem{exper} A. M. Chang et al., {\it Phys. Rev. Lett.} {\bf 76},
1695 (1996); J. A. Folk et al., {\it Phys. Rev. Lett.} {\bf 76},
1699 (1996).

\bibitem{mcdonald} S. W. McDonald, Lawrence Berkeley Lab. Report
LBL-14837 (1983).

\bibitem{earlyscar} E. J. Heller,  {\it Phys. Rev. Lett.} {\bf 53}, 1515 (1984);
E. B. Bogomolny, {\it Physica} {\bf D 31}, 169 (1988);
M. V. Berry, {\it Proc. Roy. Soc.}  {\bf A 243}, 219 (1989).

\bibitem{recentscar} 
L. Kaplan and E. J. Heller,
{\it Ann. Phys. (N. Y.)} {\bf 264}, 171 (1998); a recent review
appears in L. Kaplan, {\it Nonlinearity} {\bf 12}, R1 (1999).

\bibitem{antiscar} L. Kaplan, {\it Phys. Rev.} {\bf E 59},
5325 (1999).

\bibitem{wis} L. Kaplan, {\it Phys. Rev. Lett.} {\bf 80}, 2582 (1998).

\bibitem{ltsc} S. Tomsovic and E. J. Heller, {\it Phys. Rev. Lett.}
{\bf 67}, 664 (1991); {\it Phys. Rev.} {\bf E 47}, 282 (1993);
L. Kaplan, {\it Phys. Rev.} {\bf E 58}, 2983 (1998).

\bibitem{levelspac} U. Sivan et al.,
{\it Phys. Rev. Lett.} {\bf 77}, 1123 (1996);
S. R. Patel et al., {\it Phys. Rev. Lett.} {\bf 80}, 4522 (1998);
Y. Alhassid, Ph. Jacquod, and A. Wobst, cond-mat/9909066.

\bibitem{alh} Y. Alhassid and C. H. Lewenkopf, {\it Phys. Rev.}
{\bf B 55}, 7759 (1997).

\bibitem{scarmometer} L. Kaplan and E. J. Heller,
{\it Phys. Rev.} {\bf E 59}, 6609 (1999).

\bibitem{bb} A. B\"acker, R. Schubert and P. Stifter, {\it J. Phys.
A} {\bf 30}, 6783 (1997);
G. Tanner, {\it J. Phys.} {\bf B 30}, 2863 (1997);
P.W. O'Connor and E.J. Heller,
{\it Phys. Rev. Lett. } {\bf 61}, 2288 (1989).

\bibitem{fintemp} Y. Alhassid, M. G\"okcedag, and A. D. Stone,
{\it Phys. Rev.} {\bf B 58}, R7524 (1998).

\end{references}
\end{document}